\documentclass[twocolumn,showpacs,preprintnumbers,amsmath,amssymb]{revtex4}
\usepackage{graphicx}
\usepackage{dcolumn}
\usepackage{bm}

\begin{document}

\title{Four-dimensional CP$^1+$ U(1) lattice gauge theory 
for 3D antiferromagnets: \\
Phase structure, gauge bosons and spin liquid}

\author{Kenji Sawamura$^*$} 
 \author{Takashi Hiramatsu$^\dag$} 
\author{Katsuhiro Ozaki$^\dag$} \author{Ikuo Ichinose$^*$}
\author{Tetsuo Matsui$^\dag$}
 \affiliation{${}^*$Department of Applied Physics, Graduate School of 
Engineering, \\
Nagoya Institute of Technology, 
Nagoya, 466-8555 Japan 
}
%
\affiliation{%
${}^\dag$Department of Physics, Kinki University, 
Higashi-Osaka, 577-8502 Japan
}%

\date{\today}

\begin{abstract}
In this paper we study the lattice CP$^1$ model in 
($3+1$) dimensions
 coupled with a dynamical compact U(1) gauge field.
This model is an effective field theory of the $s={1 \over 2}$ 
antiferromagnetic Heisenberg spin model in three spatial dimensions.
By means of Monte Carlo simulations, we investigate its phase
structure. There exist the Higgs, Coulomb and confinement phases, 
and the parameter regions of these phases are clarified.
We also measure magnetization of O(3) spins, energy gap of
spin excitations, and mass of gauge boson.
Then we discuss the relationship between these three phases and 
magnetic properties of the high-$T_{\rm c}$ cuprates, in particular 
the possibility of deconfined-spinon phase.
Effect of dimer-like  spin exchange coupling 
and ring-exchange coupling is also studied. 
\end{abstract}

\maketitle
The CP$^{\rm N}$ spin model plays an important role in various 
fields of physics 
not only as a tractable field-theory model that has 
interesting phase structure, 
but also as an effective field theory for certain systems
in condensed matter physics and beyond.
In particular, the  CP$^1$ model corresponds to the 
Schwinger-boson 
representation of the $s={1 \over 2}$ antiferromagnetic (AF) 
quantum spin model, i.e., the AF Heisenberg model\cite{IM1}.
The CP$^1$ model is much more tractable than the original 
AF Heisenberg model, and its phase structure and critical
behavior have been investigated both analytically and 
numerically.
The system intrinsically contains compact U(1) gauge 
degrees of freedom, and their dynamics determines
the low-energy excitations in AF magnet.
That is, if the gauge dynamics is in the deconfined-Coulomb phase,
the low-energy excitations are the $s={1 \over 2}$ spinons.
On the other hand, the Higgs phase corresponds to the N\'eel state
with a long-range AF order, and the confinement phase is a 
valence-bond solid (VBS) state in which spin-triplet low-energy 
excitations appear.

Most of the previous studies exploring a possible  
deconfined-spin-liquid phase have considered the
two-dimensional (2D) (doped) AF Heisenberg model or
its path-integral representation, the three-dimensional
(3D) CP$^1$ model.
In these cases, the Coulomb
phase may be possible if there exist 
sufficient number of gapless matter fields that 
couple to the gauge field.
In fact, the problem whether  deconfinement in a 3D U(1) 
gauge theory is possible was once a matter of controversy,
but it seems now settled down, i.e., 
when the density of gapless excitations
in matter sector is sufficiently large, 
the deconfinement phase takes place
as a result of shielding mechanism\cite{IMO}.

In the present paper, we shall consider a higher-dimensional
version of the 3D CP$^1$ model, i.e., the 4D CP$^1$ model 
coupled with a dynamical U(1) gauge field.
This 4D CP$^1$+U(1) gauge model is viewed as an effective 
field theory of the 3D AF Heisenberg model. 
From the gauge-theoretical point of view, the deconfinement nature is 
{\em enhanced} in $(3+1)$ D case 
because the Coulomb phase exists even 
in the pure 4D U(1) gauge system that involves no matter fields
in contrast to the pure 3D U(1) gauge system that has only
confinement phase.
Therefore, it is interesting to study the phase structure of this
4D CP$^1$ gauge model. We shall first consider the 
CP$^1$ model for the 3D AF Heisenberg model with 
uniform nearest-neighbor spin coupling and then the 
CP$^1$ model for the 3D AF Heisenberg model with
 nonuniform dimer-like coupling and ring-exchange spin coupling. 

The CP$^1$ variable $z_x$ is put on 
the site $x$ of the 4D hypercubic (space-imaginary time) lattice. 
$z_x$ is a two-component complex field
satisfying the CP$^1$ constraint,
$z_x\equiv (z_{x1}, z_{x2})^t,\; \sum_{a=1}^2 |z_{xa}|^2=1$.
In path-integral, the $s={1 \over 2}$ spin operator 
$\vec{\hat{S}}_r$ at the spatial site $r$ is 
mapped to a classical O(3) spin 
$\vec{S}_x=\bar{z}_x\vec{\sigma}z_x$ 
satisfying $\vec{S}_x \cdot \vec{S}_x=1$ where $x =(r,x_0)$ 
and $\vec{\sigma}$ are the Pauli spin matrices.
The U(1) gauge field $U_{x\mu} \equiv \exp(i\theta_{x\mu})$ 
($\mu=1, \cdots, 4$ is the direction index and denotes also
the unit vector in the $\mu$-th direction)
is put on the link $(x,\mu)$ connecting sites $x$ and $x+\mu$.

The CP$^1$ field theory in the continuum is derived from
the 3D AF Heisenberg model by integrating out the half
of the CP$^1$ variables (on all the odd sites) 
by assuming a short-range AF order\cite{IM1}.
In order to study the model numerically,
we reformulate the model by putting it on the 4D 
hypercubic lattice, which has the correct continuum limit. 
Then the action $S$ of the CP$^1$ model is given as
\begin{eqnarray}
S&=&-\frac{c_1}{2}\sum_{x,\mu,a}
\Big(\bar{z}_{x+\mu,a}U_{x\mu} 
z_{xa} + \mbox{H.c.}\Big) \nonumber  \\
&&
-\frac{c_2}{2}\sum_{x,\mu<\nu}\Big(\bar{U}_{x\nu}\bar{U}_{x+\nu,\mu}
U_{x+\mu,\nu}U_{x\mu}+\mbox{H.c.}\Big),
\label{model_1}
\end{eqnarray}
where $c_1$ and $c_2$ are parameters of the model. 

Qualitative estimation of the parameters in the action (\ref{model_1})
is obtained as follows. By choosing the 
lattice spacing in the time direction $a_0$ suitably,
we obtain Eq.(\ref{model_1}) with
$c_1$ a {\it constant} 
independent of the exchange coupling $J$ of the 
AF Heisenberg model, and $c_2 = 0$\cite{coupling}. 
However, via the renormalization effect of high-momentum modes 
of spinons $z_x$, not only the $c_1$-term is renormalized but also
the $c_2$-term is generated. 

When holes are doped into an AF magnet like 
the high-$T_{\rm c}$ cuprate at half-filling, the system
is described by a canonical model like the t-J model.
The t-J model can be studied by the slave-fermion-CP$^1$ 
representation\cite{IM1}.
One can see that 
as a result of the hole doping in the short-range AF background,
the parameter $c_1$ is renormalized.
If the hopping of holons is ignored, $c_1 \rightarrow (1-\delta)c_1$
where $\delta$ is the hole concentration.
Furthermore, the short-range AF order generates attractive 
force between holes sitting on nearest-neighbor sites, and it
indicates the appearance of the superconducting phase.
When the hole-pair field $M_{x,i}\; 
(i=1,2,3: \mbox{spatial direction index})$ 
condenses, 
the $c_2$-plaquette term in Eq.(\ref{model_1}) is generated by the hopping of
quasiparticles (i.e., the gapped holons)\cite{IM1}.

\begin{figure}[t]
\includegraphics[width=7cm]{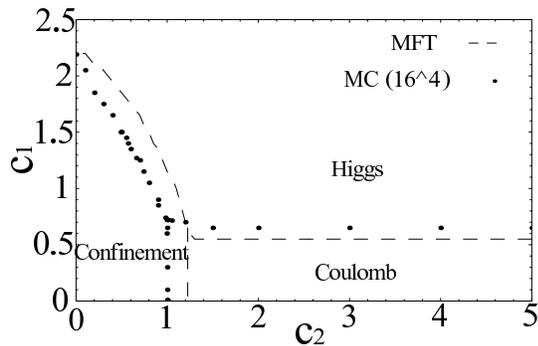}
\caption{\label{4dcp1pd}
Phase diagram of the 4D CP$^1$ model (\ref{model_1}) 
in the $c_2-c_1$ plane obtained by  
Monte Carlo simulation (circles) and by
mean field theory (dashed curves).
The Higgs-Coulomb and confinement-Coulomb phase transitions are
 of second order, whereas 
the Higgs-confinement transition is of first-order in the 
region near the tricritical point, and becomes second-order
for higher $c_1$. For example, $(c_2,c_1)=(0.96,1.27)$ is of
first order and (0.87,1.40) is of second order.
}\end{figure}


The phase diagram of the model (\ref{model_1}) 
in the $c_2-c_1$ plane has been determined
by calculating the ``internal energy" 
per site $E=\langle S \rangle/V$ and the 
``specific heat" per site
$C=\langle (S-E)^2 \rangle/V$ by means of Monte Carlo 
simulations for a lattice of size $V=L^4$ with the 
periodic boundary condition in the previous paper\cite{n-net}.
We show the result in Fig.\ref{4dcp1pd}.
In Fig.\ref{4dcp1pd}, we also show the result obtained by the 
mean-field approximation.
There are three phases in the $c_2-c_1$ plane.
We expect and verify that these three phases are {\em Higgs, 
confinement} and {\em Coulomb} phases in the gauge-theory 
terminology, respectively. The Higgs phase is nothing but 
the N\'eel state of the AF magnets.
There deconfined gapless spinons are spin-wave magnons.
The confinement phase is a spin-liquid phase in which the 
low-energy excitations are {\em spin-triplet bound states} 
of $z_{x}$.
On the other hand, the Coulomb phase corresponds to a {\em deconfined spin 
liquid} in which the low-energy excitations are  
$s={1 \over 2}$ spinons $z_x$ with a gap.
We shall see that the gauge-boson mass vanishes in the 
Coulomb phase while it remains finite in the other two phases.

In Ref.\cite{TIM1} we have studied the phase structure 
of the  CP$^1$+U(1) model (\ref{model_1}) not on the 4D lattice 
but  on the {\it  3D lattice}.
 We found that the Coulomb phase is missing.
The gauge-boson mass takes its minimum value 
just on the critical line separating 
the Higgs and confinement phases. If we include
$N_{\rm f}$-fold CP$^1$ variables, 
the gauge-boson mass  on the critical line 
vanishes for $N_{\rm f} \ge 14$\cite{TIM2}. 
The original uniform AF Heisenberg model is in the N\'eel state,
which corresponds to the Higgs phase\cite{IM1}.

\begin{figure}
\includegraphics[width=5cm]{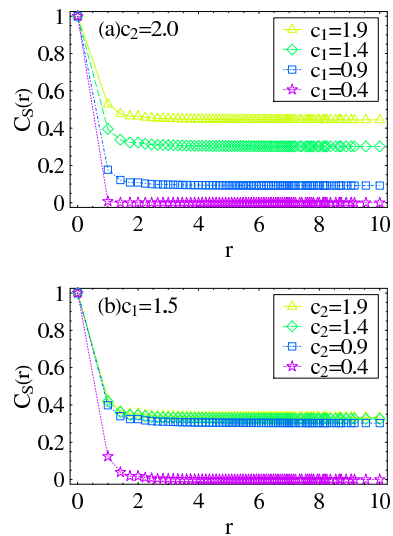}
\caption{\label{O(3)spin}
Spin-spin correlation functions $C_S(r)$
for $12^4$ lattice. 
(a) $c_2=2.0$; (b) $c_1=1.5$.
The magnetization is finite in the Higgs phase.
}\end{figure}


In Fig.\ref{O(3)spin},
we show the result of our calculations of 
the $O(3)$ spin-spin correlation functions, 
$
C_{\rm S}(r)=\langle \vec{S}_{x+r}\cdot \vec{S}_x\rangle.
$
From Fig.\ref{O(3)spin} we see that the spontaneous
magnetization,  $[C_{\rm S}(r_{\rm max})]^{1/2}\
(r_{\rm max} \equiv L)$, 
is nonvanishing only in the Higgs phase as expected.

\begin{figure}[b]
\includegraphics[width=5cm]{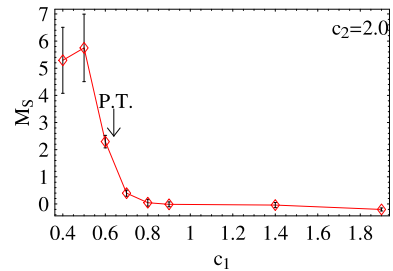}
\caption{\label{M_S}
Mass $M_{\rm S}$ of spin excitations for $c_2 = 2.0$. 
P.T. indicates the location of transition
point in Fig.\ref{4dcp1pd}.
$M_{\rm S}$ vanishes in the N\'eel state (Higgs phase).
}\end{figure}

In Fig.\ref{M_S} we show the mass gap $M_{\rm S}$
of the spin excitations  that is obtained from
 $C_{\rm S}(r)-C_{\rm S}(r_{\rm max})$.
Fig.\ref{M_S} shows that $M_{\rm S}$ is 
vanishing in the Higgs phase as expected.


\begin{figure}[t]
\hspace{-0.8cm}
\includegraphics[width=4cm]{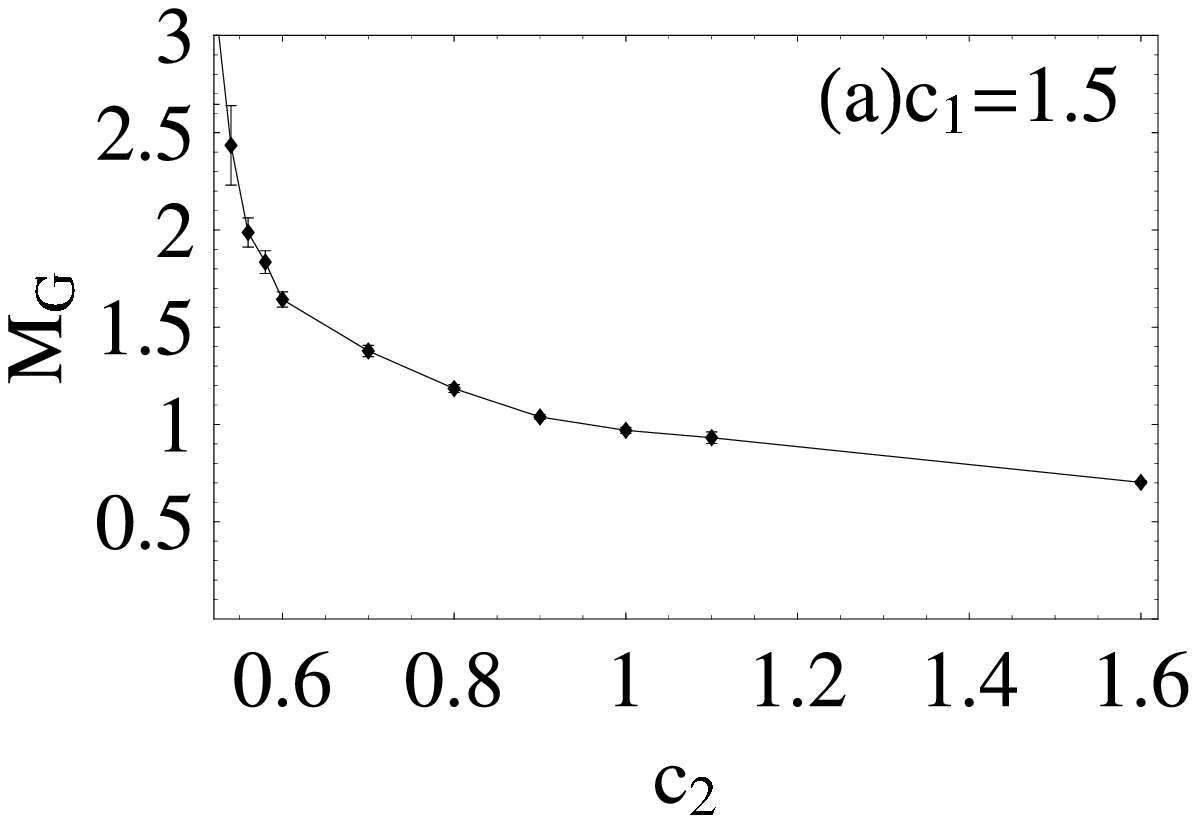}
\hspace{0.5cm}
\includegraphics[width=4cm]{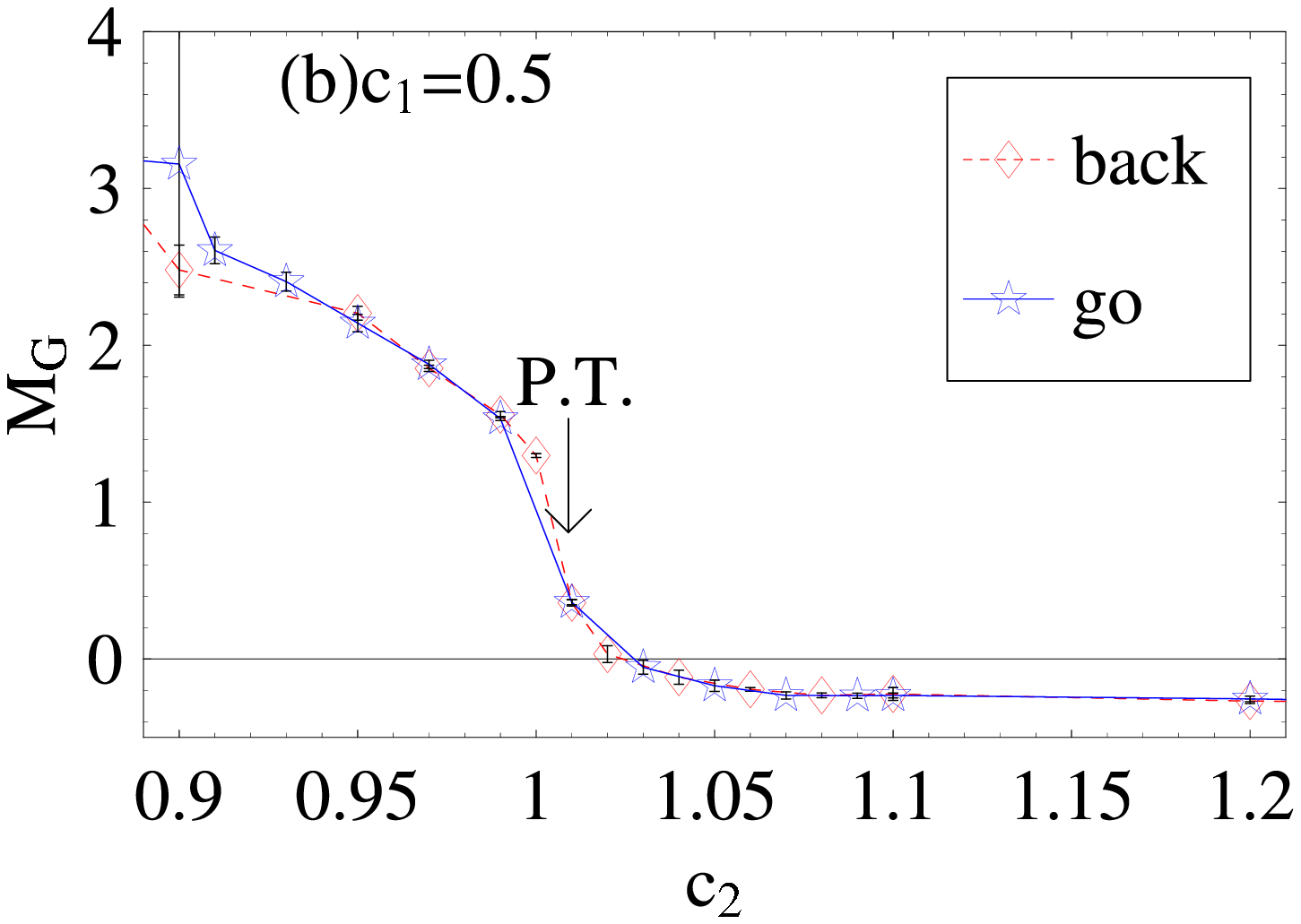}
\includegraphics[width=4.5cm]{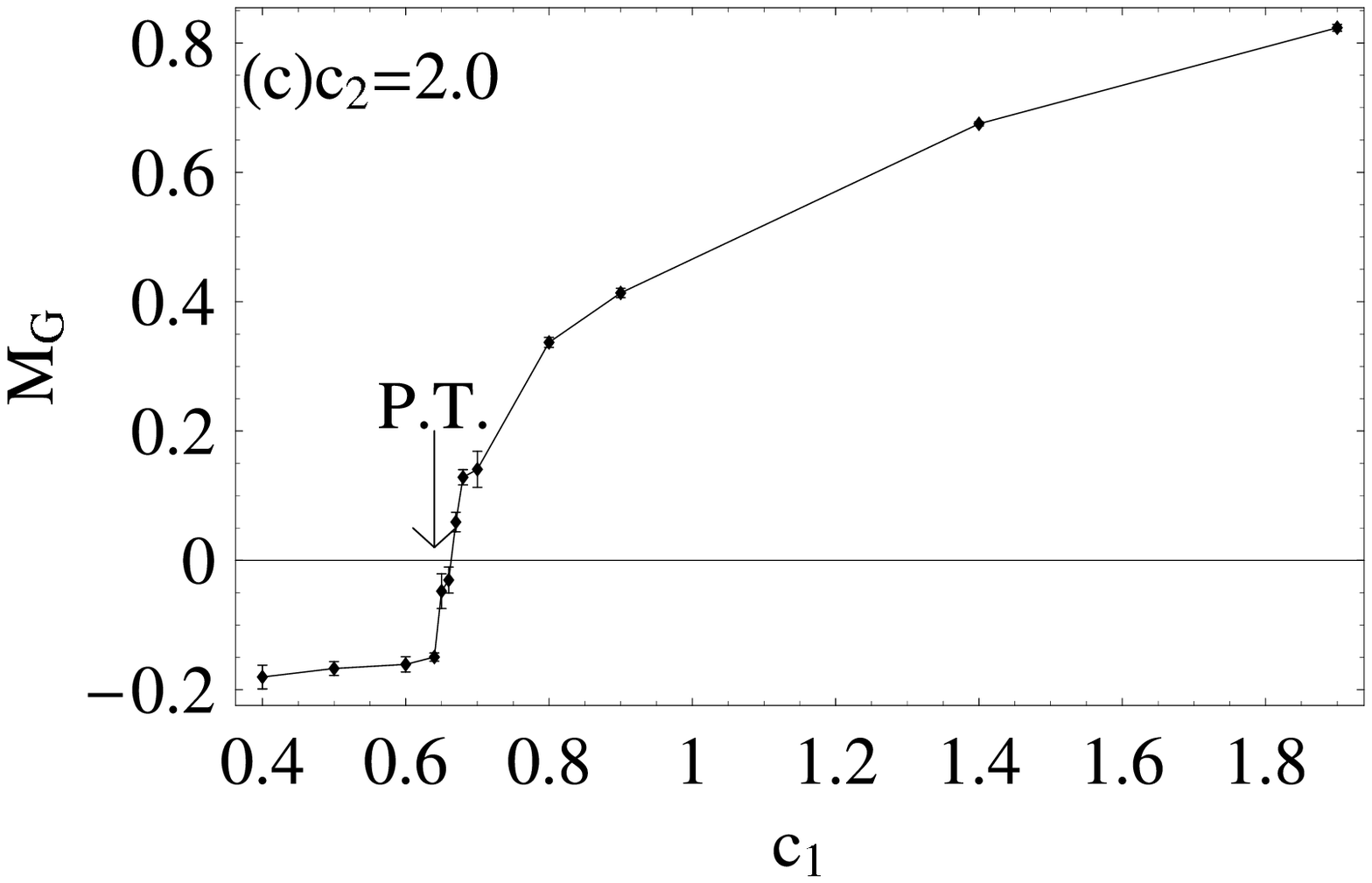}
\caption{\label{M_G}
Gauge-boson mass $M_{\rm G}$ of (\ref{gaugebosonmass}).
(a) $c_1=1.5$ (b) $c_1=0.5$ (c) $c_2=2.0.$
$M_{\rm G}$ vanishes only in the Coulomb phase.
(Negative value of $M_{\rm G}$ is a finite-size effect.
See Ref.\cite{TIM2}.)
}\end{figure}


Next, we study the dynamics of gauge bosons
by measuring the gauge-boson mass $M_{\rm G}$, which is 
calculated from the correlation function of
gauge flux,

\begin{figure}[b]
\includegraphics[width=3.5cm]{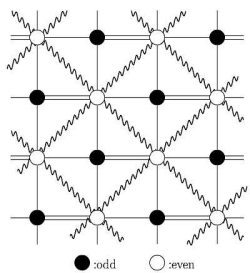}
\caption{\label{fig:model2}
AF Heisenberg model with dimer and ring couplings.
The double-line bonds have larger exchange than that of the
single-line bond. The wavy lines represent ring coupling
among four spins on each square of even-sites.
}
\end{figure}


\begin{equation}
O_{\mu\nu}(x)=\sum_{\lambda, \eta}\epsilon_{\mu\nu\lambda\eta}
{\rm Im}\bar{U}_{x\lambda}\bar{U}_{x+\lambda,\eta}
U_{x+\eta,\lambda}U_{x\eta},
\label{O1}
\end{equation}
where $\epsilon_{\mu\nu\lambda\eta}$ is the totally 
antisymmetric tensor.
To estimate $M_{\rm G}$ precisely, we introduced the 
Fourier-transform $\tilde{O}$ and measured 
its correlation,
\begin{eqnarray}
\tilde{O}(x_\mu,x_\nu)=\sum_{x_\lambda,x_\eta}
O_{\mu\nu}(x)e^{ip_\lambda x_\lambda+ip_\eta x_\eta},
\nonumber\\
\label{O2}
D_G(y_\mu,y_\nu)={1 \over L^4}\sum_{x_\mu,x_\nu}
\langle \tilde{O}(x_\mu,x_\nu)
\bar{\tilde{O}}(x_\mu+y_\mu,x_\nu+y_\nu)\rangle,
\label{O3}
\end{eqnarray}
with setting $p_\lambda=p_\eta=1/L$ where $L$ 
is the system size.
We fit the data in the following form\cite{MG};
\begin{equation}
D_G(y_\mu,y_\nu) \propto \exp\left(
-\sqrt{p^2_\mu+p^2_\nu+M^2_{\rm G}}
 \sqrt{y^2_\mu+y^2_\nu}\right).
 \label{gaugebosonmass}
\end{equation}
In Fig.\ref{M_G}, we show $M_{\rm G}$.
We can see that $M_{\rm G}$ is 
vanishing in the 
Coulomb phase as it should be, whereas it is finite in the other phases.

The above measurements confirm the phase structure 
of the model (\ref{model_1}) given in Fig.\ref{4dcp1pd}.
From this result, we can discuss how the magnetic properties of the
doped AF magnets with strong three-dimensionality changes under doping
at zero temperature.
Undoped and lightly doped uniform AF magnets in 3D has the N\'eel order 
and is in the Higgs phase.
As $\delta$ is increases, the system loses the N\'eel order and enters
the confinement or Coulomb phase depending on the mobility of doped holes.
In the framework of the t-J model, this problem is under study and 
the result will be published in near future\cite{IMS}.

\begin{figure}[t]
\includegraphics[width=5cm]{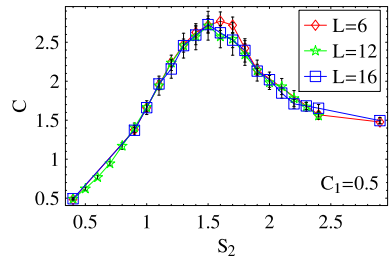}
\caption{\label{C_(3+1)}
Specific heat of the 3D plaquette model (\ref{model_2}) 
for $c_1=0.5$ as a function of $s_2$. 
There is no system-size dependence in $C$, indicating 
a crossover near the round peak.
}\end{figure}


\begin{figure}[b]
\includegraphics[width=5.0cm]{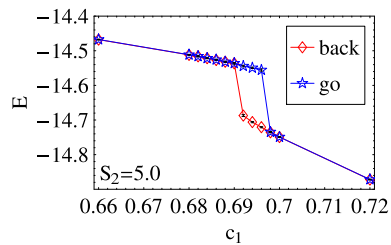}
\caption{\label{E_(3+1)}
Internal energy of the 3D plaquette model (\ref{model_2}) 
for $s_2=5.0$. It shows a hysteresis curve.
}\end{figure}


In the rest of this paper,
we shall consider the CP$^1$ gauge model with the 3D 
spatial gauge-plaquette term. The action $S'$ 
of this 3D plaquette model is given by
\begin{eqnarray}
S'&=&-\frac{c_1}{2}\sum_{x,\mu,a}\Big(\bar{z}_{x+\mu,a}U_{x\mu} 
z_{xa} + \mbox{H.c.}\Big) \nonumber  \\
&&
-\frac{s_2}{2}\sum_{x,i,j=1(i<j)}^3
\Big(\bar{U}_{xj}\bar{U}_{x+j,i}
U_{x+i,j}U_{xi}+\mbox{H.c.}\Big).
\label{model_2}
\end{eqnarray}
In contrast to the $c_2$ term of the
4D plaquette  model (\ref{model_1}), 
the $s_2$-term in Eq.(\ref{model_2})
contains only the {\em spatial-plaquette} terms
and no space-time plaquette terms. 
The 3D plaquette model (\ref{model_2}) is an effective low-energy model
of {\em nonuniform} AF Heisenberg model with dimer-like 
couplings and ring-exchange couplings.
The parameter $c_1$ decreases as the dimer coupling 
increases\cite{dimer} and the $s_2$ term
corresponds to  ring coupling
like $(\vec{S}_i\cdot \vec{S}_j)(\vec{S}_k\cdot \vec{S}_l)$
where $(i,j,k,l)$ denote even sites forming a square 
in even sublattice. (See Fig.\ref{fig:model2}.) 
Thus the bare value of $s_2$ is finite in contrast to
the 4D plaquette model (\ref{model_1}) for the symmetric 
Heisenberg model where the bare value of $c_2=0$.

\begin{figure}[t]
\includegraphics[width=5cm]{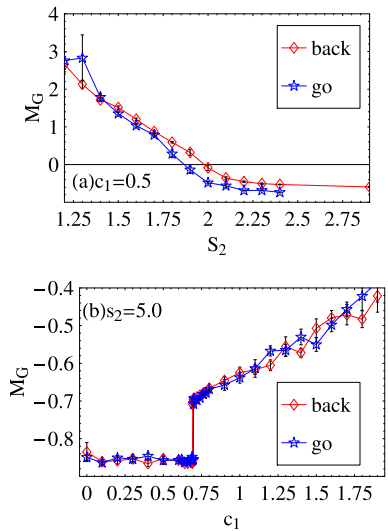}
\caption{\label{M_(3+1)_2}
Gauge boson mass $M_{\rm G}$ in the model (\ref{model_2}). 
(a) $c_1=0.5$; (b) $s_2=5.0$\cite{FN}.
}
\end{figure}


It is interesting to see if the Coulomb phase (deconfined 
spin-liquid phase), which exists in the model (\ref{model_1}), 
survives in the present model.
Therefore we investigated the phase structure of the model 
(\ref{model_2}) first by calculating
the internal energy and the specific heat.
Some of our results are shown in Figs.\ref{C_(3+1)} and \ref{E_(3+1)}.
From these calculations, we conclude that the model $S'$
of (\ref{model_2})
has similar phase structure to that of $S$ of (\ref{model_1}).
But the second-order confinement-Coulomb phase transition 
in the model $S$ becomes a {\em crossover} in $S'$ 
as Fig.\ref{C_(3+1)} 
shows, i.e., the specific heat has a rather smooth peak 
that has {\em no system-size dependence}.
We think that this crossover is similar to that in 
the 3D Abelian gauge-Higgs model. 
Furthermore, the phase transitions at $c_1 \simeq 0.7$
in the region of $s_2 \stackrel{>}{\sim}1.5$ are of 
{\em first order} as indicated in Fig.\ref{E_(3+1)}.

In order to verify the above conclusion, we calculated 
$M_{\rm G}$, which is shown in Fig.\ref{M_(3+1)_2}.
It is obvious that along with $c_1=0.5$ 
$M_{\rm G}$ decreases smoothly 
as $s_2$ is increased crossing the crossover line at 
$s_2 \simeq 1.5$.
This behavior is very close to that in the Abelian gauge-Higgs 
system observed in Ref.\cite{TIM2}.
On the other hand, along the line $s_2=5.0$ 
$M_{\rm G}$ exhibits a sharp
discontinuity. Its behavior verifies 
the first-order phase transition
at $c_1 \simeq 0.7$\cite{FN}.

Then we draw the following conclusions;\\
1. In contrast to the 4D plaquette model $S$, the 
3D plaquette model $S'$
does not support the Coulomb phase, i.e., the 
deconfined spin-liquid phase is not realized in the 
nonuniform AF Heisenberg model\cite{AHM,AFHM}.\\
2. The N\'eel-VBS phase transition caused by the 
ring-exchange coupling is of first-order. 
Recent numerical study on some related AF Heisenberg 
models gives similar conclusion\cite{AFHM}.

In summary, we studied the 4D CP$^{\rm 1}$ model 
coupled with 4D or 3D U(1) gauge field and revealed its 
phase structure. We discussed some physical implications 
for various 3D AF Heisenberg magnets, i.e., 
with and without doping and with uniform or nonuniform coupling
in gauge-theoretical viewpoints.



\begin{references} 


\bibitem{IM1}I.Ichinose and T.Matsui, 
Phys.Rev.{\bf B45}, 9976(1992); \\
H.Yamamoto, G.Tatara, I.Ichinose, and T.Matsui, \\
Phys.Rev.{\bf B44},7654(1991).
Strictly speaking, there appears an imaginary phase-term in the
effective action. We neglect it here expecting that it does not
affect the phase structure.
See, for example, S.Kragset, E.Sm\o rgrav, \\
J.Hove, F.S.Nogueira, and A.Sudb\o,
Phy.Rev.Lett. {\bf 97},\\
247201(2006).

\bibitem{IMO}I.Ichinose and T.Matsui, 
Phys.Rev.Lett.{\bf 86}, 942(2001) ;\\
I.Ichinose, T.Matsui, and M.Onoda, Phys.Rev.{\bf B64}, \\
104516(2001) and references cited therein;
G.Arakawa, I.Ichinose, T.Matsui, K.Sakakibara, 
Phys. Rev. Lett.{\bf 94},\\
211601(2005). 

\bibitem{coupling}
Derivation of the effective CP$^1$ model
in the continuum limit is discussed in detail in the first paper of
Ref.\cite{IM1}.

\bibitem{n-net}
T.Hiramatsu and T.Matsui, 
in preparation.


\bibitem{TIM1}S.Takashima, I.Ichinose, and T.Matsui,
Phys.Rev.{\bf B72}, \\
075112(2005).

\bibitem{TIM2}S.Takashima, I.Ichinose, and T.Matsui,
Phys.Rev.{\bf B73}, \\
075119(2006).


\bibitem{MG}More precisely, we fit the correlator as
$
D_G(y_\mu,y_\nu) \propto \exp(-\lambda \sqrt{y^2_\mu+y^2_\nu}),$
by adjusting a fitting parameter $\lambda$. 
Then we define $M_{\rm G}$ as
$
M_{\rm G}={\rm sgn} (\lambda^2 -p^2_\mu-p^2_\nu)
\sqrt{|\lambda^2 -p^2_\mu+p^2_\nu|}.
$
It has been verified that negative values of $M_{\rm G}$ are 
finite-size effect.
See Refs.\cite{TIM1,TIM2}.


\bibitem{IMS}I.Ichinose, T.Matsui, and K.Sakakibara,
work in \\
progress.

\bibitem{dimer}D.Yoshioka, G.Arakawa, I.Ichinose, and T.Matsui,\\
Phys.Rev.{\bf B70}, 174407(2004).

\bibitem{FN}Negative values of $M_{\rm G}$ in Fig.\ref{M_(3+1)_2}
come from the definition of $M_{\rm G}$.
The plaquette term with fairly large coefficient ($s_2=5.0$) severely 
suppresses fluctuations of the gauge field in the present case.

\bibitem{AHM}Nonetheless, for the Abelian gauge-Higgs model in 3D, 
it was suggested that the across the crossover line 
certain nonlocal physical quantities like flux-line 
density exhibit typical behavior of second-order phase 
transition. 
See S.Wenzel, E.Bittner, W.Janke, and A.M.J.Schakel, 
arXiv:0708.0903.
We expect similar behavior for the present model $S'$.

\bibitem{AFHM}See,  for example, K.S.D.Beach and A.W.Sandvik,\\
Phys.Rev.Lett.{\bf 99}, 047202(2007).


\end{references}
\end{document}